\documentclass[%
mathleft,%
]{an}
\usepackage{graphicx}
\usepackage{times}
\usepackage{amsmath}
\usepackage{hyperref}

\begin{document}

\Pagespan{1}{}
\Yearpublication{2015}%
\Yearsubmission{2014}%
\Month{1}%
\Volume{999}%
\Issue{92}%

\title{CoRoT\thanks{The CoRoT space mission, launched on 2006 December 27, has been developed and is operated by CNES, with the contribution of Austria, Belgium, Brazil, ESA (RSSD and Science Programme), Germany and Spain.} Data Reduction By Example}

\author{J. Weingrill\inst{1}\mail{jweingrill@aip.de}}

\titlerunning{CoRoT Data Reduction By Example}
\authorrunning{J. Weingrill}
\institute{
Leibniz-Institut f\"{ur} Astrophysik (AIP), An der Sternwarte 16, 
D-14482 Potsdam, Germany}

\received{\today}
\accepted{XXXX}
\publonline{later}


\abstract{%
Data reduction techniques published so far for the CoRoT N2 data product were targeted primarily on the detection of extrasolar planets.
Since the whole dataset has been released, specific algorithms are required to process the lightcurves from CoRoT correctly.  
Though only unflagged datapoints must be chosen for scientific processing, some flags might be reconsidered. The reduction of data along with improving the signal-to-noise ratio can be achieved by applying a one dimensional drizzle algorithm. Gaps can be filled by linear interpolated data without harming the frequency spectrum. Magnitudes derived from the CoRoT color channels might be used to derive additional information about the targets. Depending on the needs, various filters in the frequency domain remove either the red noise background or high frequency noise.
The autocorrelation function or the least squares periodogram are appropriate methods to identify periodic signals.
The methods described here are not strictly limited to CoRoT data but may also be applied on Kepler data or the upcoming Plato mission.}

\maketitle


\section{Introduction}



Data reduction techniques specific to the transit detection can be found in various papers, e.g. (Carone et al. 2012, Bonomo et al. 2012, Grziwa et al. 2012), but so far no general description for data reduction of the CoRoT N2 data product is available. 

Attempts have been made in the past to deal with partial problems like the minimization of systematic effects by Mazeh et al. (2009) or the correction of jumps in chromatic lightcurves by Mislis et al. (2010). But no complete `toolbox' of algorithms and routines for the data reduction of the CoRoT dataset has been published yet. The methods described here are not strictly limited to CoRoT data, but may also be applied on \textit{Kepler} (Borucki et al., 2010) data or the upcoming \textsc{Plato}~2.0 mission (Rauer et al., 2014). 

The CoRoT satellite is an advanced high precision photometric instrument devoted to the detection of transiting extrasolar planets and asteroseismology (Baglin et al. 2006). A description of the satellites performance and characteristics can be found in the paper by Auvergne et al. (2009). This paper mainly focuses on the exo-channel of the CoRoT data product. Its noise characteristics are extensively analysed by Aigrain et al. (2009). 

Several science-applications require an additional processing beyond the given N2~data-product to extract the information. This data-product, which is published to the end-user only contains corrections that can be deduced from a deterministic model like e.g. the satellite's orbit. The corrections applied are for electronic linearity, interference, outliers, background and jitter (Auvergne et al., 2009). Other variations either of systematic or random origin have to be removed by the user. This data-product is unfiltered by intention to allow for different scientific processing. A prominent example is the search for transit-like signals which require the stellar activity and the instrumental noise to be removed in contrast to the identification of stellar magnetic activity as stellar spots for instance where transit signals dilute the lightcurve. 

Data from CoRoT runs has been made available for download at the IAS data centre\footnote{\url{http://idoc-corot.ias.u-psud.fr/}}. Supplementary information can be accessed through ExoDat\footnote{\url{http://cesam.oamp.fr/exodat/}} (Meunier et al. 2007; Deleuil et al. 2009). It is highly recommended to use VOTools like e.g. Topcat\footnote{\url{http://www.star.bris.ac.uk/~mbt/topcat/}} or Aladin\footnote{\url{http://aladin.u-strasbg.fr/}} to access the ExoDat database. Most auxiliary information in the FITS-files has been assembled by using the 2MASS (Skrutskie et al. 2006), NOMAD (Zacharias et al. 2005) and DENIS (Vauglin et al. 1999) catalogue. The given spectral type, luminosity class, or effective temperature represent only preliminary information and should be compared with other source catalogues for correctness. The given H-, J- and K-magnitudes are rather precise and can be used for a rough stellar spectral classification. 

\begin{figure}[ht]
\includegraphics[width=\columnwidth]{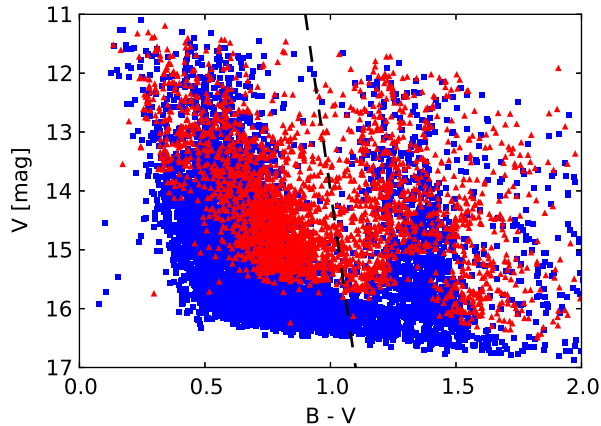}
\caption{Color-magnitude diagram of stars from LRa01. Dwarf stars are represented as squares, giants as triangles. A possible criterion for the separation of luminosity classes is indicated by a dashed line.\label{fig:cmd}}
\end{figure}

Though the luminosity class estimate is given for most of the stars in the FITS-header information, the position in the color-magnitude diagram should be used as a reference to roughly discriminate between the dwarf and giant population. Usually the information from ExoDat succeeds the information from the FITS-headers, requiring the user to perform the update of relevant data by himself. An example of possible (mis-)identification of luminosity classes of stars in the run LRa01 (Carone et al., 2012) is shown in Figure~\ref{fig:cmd}. 
For parameters resulting in astrophysical parameters like e.g. $B-V$, one should use VizieR\footnote{\url{http://vizier.u-strasbg.fr/viz-bin/VizieR}} to get up-to-date measurements.

We are only discussing the reduction of the exoplanet-channel here since the requirements for the asteroseismology are very different to preserve the precious frequency information. However, most of the algorithms and filter techniques presented here could in principle also be applied on the asteroseismic-channel. More detailed information on the CoRoT's seismology programme can be found in Michel et al. (2006).

The CoRoT observation started with the initial field IRa01 (Carpano et al., 2009) in the galactic anti-center direction towards the constellation Monoceros. The next pointing was a so-called short run SRc01 (Erikson et al., 2012) in the galactic center-direction between the constellations Aquila and Ophiuchus. The two fields in the constellations are called \textit{eyes of CoRoT} and are necessary to avoid contamination by sunlight and the eclipse of the satellite by the earth (Baglin et al, 2000). Some areas within the field of view of CoRoT have been visited twice. The CoRoT observation runs lasted between 20 days for the short ones up to 150 days for the long ones, which are listed in table~\ref{tab:runlengths}. The detector E1 failed between the observation runs LRa02 and LRc03. (Moutou et al., 2013)

\begin{table}
 \centering
\caption{List of CoRoT runs with their approximate run length sorted by date of observation start. The run LRc09 was not public available at the date of publication.}
\label{tab:runlengths}
\begin{tabular}{lr||lr}
run & duration & run & duration\\
code &  (days) & code &  (days)\\
\hline
 IRa01 &  57.75 &  SRa03 & 24.29\\
 SRc01 &  25.55 &  LRc05 &  87.26\\
 LRc01 & 152.02 &  LRc06 &  77.41\\
 LRa01 & 131.46 &  LRa04 &  77.58\\
 SRa01 &  23.41 &  LRa05 &  90.49\\
 LRc02 & 145.00 &  LRc07 &  81.21\\
 SRc02 &  20.88 &  SRc03 &   3.46\\
 SRa02 &  31.74 &  LRc08 &  83.56\\
 LRa02 & 114.65 &  SRa04 &  52.25\\
 LRc03 &  89.19 &  SRa05 &  38.68\\
 LRc04 &  84.15 &  LRa06 &  76.62\\
 LRa03 & 148.31 &  LRc09 &     --\\

 \hline
\end{tabular}
\end{table}

Each target is assigned a unique identification number, the so-called CoRoTID. Additionally the targets can also identified by the combination of \textsc{Run\_ID}, \textsc{HlfCCD\_ID} and \textsc{Win\_ID}, which is also unique for each CoRoT target (Carone et al., 2012). To verify the existence of multiple observations for one target, the CoRoTIDs have to be queried in the ExoDat database. The user has to download the whole run and concatenate the observations based on the time-stamp. A proper cross-identification of targets should be done using their coordinates. Individual targets can be downloaded on ExoDat as FITS files or by using VO-tools.

Except for the runs SRa01 (pipeline version~0.9 and N2-version~2.1) and SRa05 (pipeline ver.\,2.3, N2-ver.\,3.0) which have erroneous coordinates in the FITS-headers, all other targets can be cross-matched using their coordinates, which originate from the USNO-B1 catalogue (Monet et al. 2003) or PPMXL catalogue (R\"{o}ser, Demleitner \& Schilbach 2010). Errors in position of CoRoT targets should always be within one arcsecond. The user should keep in mind that the point-spread-function (PSF) is several arcseconds wide and that the variability might not originate from the CoRoT target, but a background star. A nice demonstration of topic is given in the paper by Papar\'{o} et al. (2011). The contamination factor from ExoDat or the FITS-header is a good proxy for the identification of diluting background targets in the PSF. Otherwise the \textsc{WinDescriptor} image must be checked for other pointsources within the PSF.

Details for some individual CoRoT runs can be found in Cabrera et al. (2009), Carone et al. (2012), Cavarroc et al, (2011) and Erikson et al. (2012).


\section{The CoRoT N2 Data Product}
\label{sec:dataproduct}
The basis for this section is the technical description for the N2 data product (Baudin et al., 2012). The lightcurves for each run are provided as gzipped tar-archives containing the \textsc{Fits}-files for each individual target. The archives are grouped by monochromatic-, chromatic-, asteroseismic- and imagette-targets as well as their \textsc{Windescriptor}-files. The imagette-archives also contain their \textsc{Windescriptor}-files.

The header keywords are subject to change in each new version of the N2-pipeline. The \textsc{Fits}-files generally follow the standard described in Hanisch et al. (2001), but contain some non-standard values for the fields. Where some values are simply left empty like e.g. the \textsc{VarClas} cards, others like unknown magnitudes are set with '-99.0000' or set with non-standard values like 'null' or 'NaN' which are not always understood by libraries used for reading the files.


\subsection{Epoch of measurements}
\label{sec:epoch}
For high precision epoch determination the \textsc{DateHel} field in the FITS-file must be used. Each 512 second exposure epoch must undergo a time-correction that is 224 seconds resulting from $t' = t + (256-32)/86400$, where $t$ is the original epoch in the file and $t'$ is the corrected epoch. For alarm-mode candidates with mixed cadence, the 32 second cadence data must not be corrected, but the 512 second data must undergo this correction. Lightcurves only taken in 32 seconds cadence mode are also exempt from this correction.

\textsc{DateJD} describes the CoRoT Julian day where day zero is defined as 1 Jan 2000 12:00:00.0 equivalent to Julian day 2\,451\,545.0 in the reference frame of the CoRoT satellite (Auvergne et al., 2009).


\subsection{Flux measurements}
CoRoT provides different products as shown in Table \ref{tab:modes} depending on the target magnitude. Some bright stars have been observed in the so-called  \textit{Imagette} mode, which should provide a high signal-to-noise ratio.
\begin{table}
\caption{Grid of different channels and modes}
\label{tab:modes}
\begin{tabular}{lll}\hline
mean values & normal mode & Imagette mode\\ 
\hline
LC\_MEAN & WHITEFLUX & WHITEFLUX\_IMAG \\
LC\_MEANR & REDFLUX & REDFLUX\_IMAG\\
LC\_MEANG & GREENFLUX & GREENFLUX\_IMAG\\
LC\_MEANB & BLUEFLUX & BLUEFLUX\_IMAG\\
\hline
\end{tabular}
\end{table}
In `CoRoT-terms' the name \textit{Imagette} also describes the aperture for the photometry (Baudin et al., 2012; Drummond et al., 2006). Each aperture is divided in three sections for the blue, red and green channel, representing the spectral output of the prism mounted in front of the charge coupled devices (CCDs) (Rouan et al., 1999; Auvergne et al., 2009). There are 256~different templates for apertures in use since the PSF varies along field of view (Barge et al., 2006). 

Usually the three colours should be at the same order of magnitude. Spacecraft jitter and relativistic aberration might cause the flux to be shifted from one channel to another. The flux distribution in the three different channels depends on the type of Imagette, position on the chip and the spectral type of the star (Auvergne et al., 2009). Several (SRa01, SRc01, SRa02 and SRc02) runs have rounded magnitudes in their headers which badly correlate to the red flux.

The CoRoT colours are still reliable enough to identify a blended binary (Carone et al., 2012) or a change in the stellar effective temperature, e.g. to differentiate between pulsation or rotation (Weingrill, 2011).  The scientific use of CoRoT's colors is still controversial (Borsa \& Poretti, 2012; Auvergne et al., 2009). For each run the flux values in the three channels (REDFLUX, GREENFLUX, BBLUEFLUX) can be transformed into CoRoT colors ($R_C$, $V_C$, $B_C$) using an idealised transformation matrix
\begin{equation}
\left(\begin{array}{c}
R_C \\ 
V_C \\ 
B_C
\end{array} \right)=\left(\begin{array}{c}
R_0 \\ 
V_0 \\ 
B_0
\end{array} \right) - 2.512\log_{10}\left(\begin{array}{ccc}
r & 0 & 0 \\ 
0 & g & 0 \\ 
0 & 0 & b
\end{array} \right),
\end{equation}
where $R$, $V$, $B$ are the respective magnitudes given in the FITS file, $r$, $g$, $b$ are the mean flux values (LC\_MEANR, LC\_MEANG and LC\_MEANB) and $R_0$, $V_0$, $B_0$ are the magnitude zero points. This is of cause the ideal case, where we assume that we have to deal with discrete filter bands. Since the color information is spread across the PSF, the color channels are polluted. For the CoRoT run LRa01 as an example we obtain
\begin{multline}\label{eqn:translra01}
\left(\begin{array}{c}
R_C \\ 
V_C \\ 
B_C
\end{array} \right)=\left(\begin{array}{c}
25.197 \\ 
25.392 \\ 
25.794
\end{array} \right)\\ 
- 2.512\log_{10}\left(\left[\begin{array}{ccc}
0.984 & -0.262 & 0.220 \\ 
0.398 & 0.155 & 0.412 \\ 
0.115 & 0.360 & 0.497
\end{array} \right]\cdot \left(\begin{array}{c}
r\\ g\\ b\\ \end{array}\right)
\right),
\end{multline}
to get the CoRoT magnitudes $R_C$, $V_C$, $B_C$. The transformation matrix in eqn.~\ref{eqn:translra01} demonstrates that most of the V flux is distributed among the red and blue channels. The transformation can be calculated for all runs, where the respective BVR-magnitudes are known. This excludes SRc02, SRa02 and LRc03 where the photometric data is incomplete. Other runs like IRa01, SRc01, LRa04 and SRc03 give unrealistic results due to e.g. negative correlation between colors and Johnson magnitudes.

\begin{figure}[ht]
\includegraphics[width=\columnwidth]{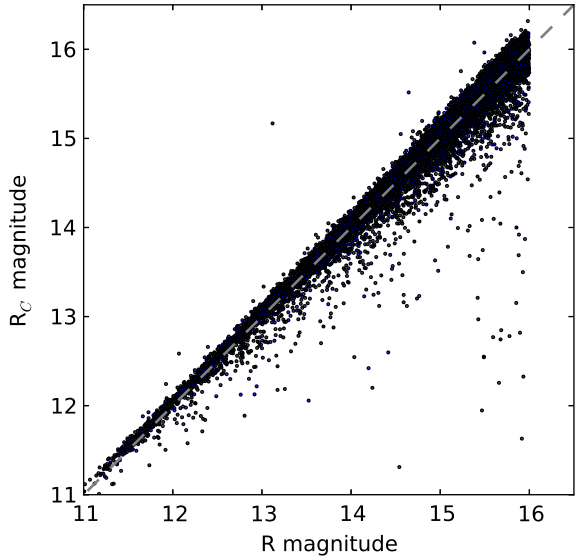}
\caption{Each dot represents a chromatic target in the run LRa01 with its given R magnitude versus CoRoT R$_C$ magnitude. The linear correlation is evident.\label{fig:magmean}}
\end{figure}
For monochromatic targets the R magnitude matches the median flux value as shown in Figure~\ref{fig:magmean}.

\begin{figure}[ht]
\includegraphics[width=\columnwidth]{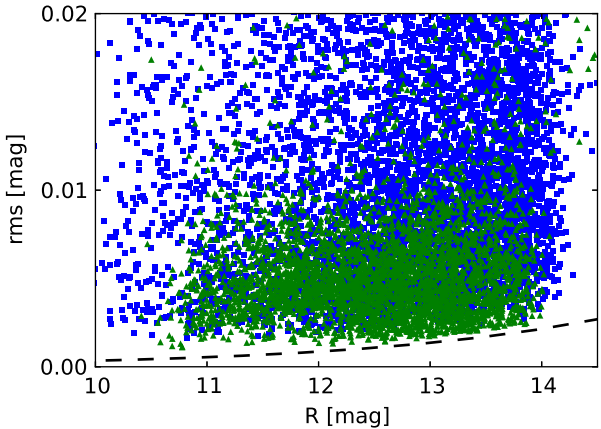}
\caption{Magnitude versus rms for each individual star in the initial run IRa01 (triangles) and the long run LRa01 (squares). The photon noise limit is indicated by a dashed line at the bottom.\label{fig:magrms}}
\end{figure}

The root-mean-square (rms) value given in the FITS-header is a preliminary indicator for the variability of the lightcurve and is mostly governed by photon noise. The rms is not corrected for any instrumental trend in the lightcurve, so for example relativistic aberration amplifies the rms value artificially. Other sources that might affect this value are the stellar activity, eclipse-signals and instrumental noise especially for faint objects. The typical noise characteristics can be seen in Figure~\ref{fig:magrms}, which compares the given rms of each lightcurve (represented by a dot) between the first run IRa01 (Carpano et al., 2009) and LRa01 (Carone et al., 2012).

The global rms value is different for each run and is affected by the ageing of the instrument (Moutou et al., 2013) as well as other sources that affect the performance of the CCDs, e.g. cosmic ray hits or changing detector temperature. The rms value given in Table~\ref{tab:runrms} is governed by the length of the run as well as by selection effects like the stellar population in the field, number of targets in the CCDs and the essential fraction of variable stars.

\begin{table}
 \centering
\caption{This table shows the overall relative rms values taken from the N2 dataproduct for each run on each CCD. The rms values were taken from monochromatic targets and averaged for each CCD and each run. The four runs LRc06, LRa04, SRc03 and LRc08 do not contain rms values in the official dataproduct.}
\label{tab:runrms}
\begin{tabular}{lccl}
start date      & run & CCD & rms  \\
\hline
2007-02-03 13:05:53 & IRa01    & E1    & 0.0075 \\
2007-02-06 13:35:47 & IRa01    & E2    & 0.0080 \\
2007-04-13 18:00:30 & SRc01    & E1    & 0.0055 \\
2007-04-13 18:02:06 & SRc01    & E2    & 0.0053 \\
2007-05-16 06:00:50 & LRc01    & E1    & 0.0126 \\
2007-05-16 06:02:26 & LRc01    & E2    & 0.0133 \\
2007-10-23 22:30:35 & LRa01    & E1    & 0.0152 \\
2007-10-23 22:30:35 & LRa01    & E2    & 0.0196 \\
2008-03-07 21:50:33 & SRa01    & E2    & 0.0079 \\
2008-03-07 21:50:33 & SRa01    & E1    & 0.0111 \\
2008-04-15 23:10:48 & LRc02    & E2    & 0.0204 \\
2008-04-15 23:10:48 & LRc02    & E1    & 0.0186 \\
2008-09-15 09:33:11 & SRc02    & E2    & 0.0415 \\
2008-09-15 09:33:11 & SRc02    & E1    & 0.0063 \\
2008-10-11 14:30:35 & SRa02    & E2    & 0.0152 \\
2008-10-11 14:30:35 & SRa02    & E1    & 0.0137 \\
2008-11-16 19:02:24 & LRa02    & E1    & 0.0126 \\
2008-11-16 19:02:24 & LRa02    & E2    & 0.0132 \\
2009-04-03 22:00:30 & LRc03    & E2    & 0.0175 \\
2009-07-07 04:45:28 & LRc04    & E2    & 0.0099 \\
2009-10-03 22:31:49 & LRa03    & E2    & 0.0172 \\
2010-03-05 00:15:25 & SRa03    & E2    & 0.0074 \\
2010-04-08 22:30:49 & LRc05    & E2    & 0.0098 \\
2010-07-08 20:45:34 & LRc06    & E2    & 0.0114 \\
2010-07-09 00:00:14 & LRc06    & E2    &  --    \\
2010-09-29 14:00:53 & LRa04    & E2    &  --    \\
2010-12-21 18:00:26 & LRa05    & E2    & 0.0154 \\
2011-04-08 14:15:49 & LRc07    & E2    & 0.0106 \\
2011-07-01 18:30:50 & SRc03    & E2    &  --    \\
2011-07-08 15:30:29 & LRc08    & E2    &  --    \\
2011-07-08 15:38:29 & LRc08    & E2    & 0.0117 \\
2011-10-07 00:10:47 & SRa04    & E2    & 0.0127 \\
2011-12-01 18:15:27 & SRa05    & E2    & 0.0344 \\
2012-01-12 18:30:55 & LRa06    & E2    & 0.0154 \\
\end{tabular}
\end{table}

To convert the rms flux to magnitude units one easily calculates $m = -2.512\log_{10}(1-r/\bar{f})$,
where $m$ is the variation in magnitudes, $r$ is the rms of the flux and $\bar{f}$ is the mean flux. Usually the rms or the variation is calculated an basis of a given time-scale, like 5 hours.
\begin{figure}[ht]
\includegraphics[width=\columnwidth]{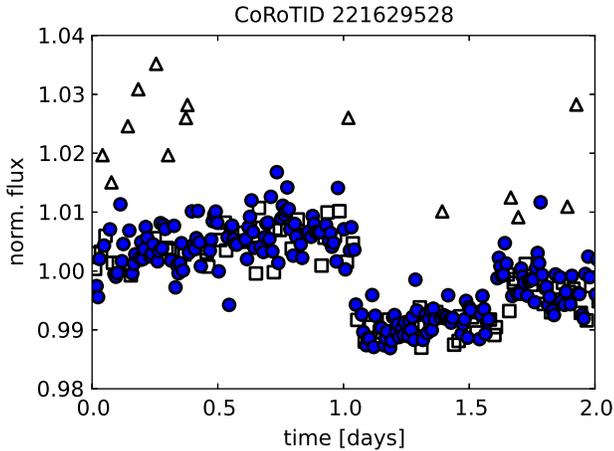}
\caption{This exemplary lightcurve demonstrates the typical distribution of data points with different flags. Valid data points are shown with filled circles, flagged data with even status (boxes) and odd ones (triangles).\label{fig:status}}
\end{figure}

All valid data points can be selected by taking the rows, where the \textsc{Status} flag is zero, which is usually more then 66\% of the data. If the number of data points is of critical issue then one might consider including flagged points where the lowest bit in the status flag is not set. This should increase the number of usable data to a maximum of 95\% (see Figure~\ref{fig:status}). The usage of flagged data still might introduce spurious artefacts or frequencies. 
All non-zero status values indicate artificial changes of the flux mostly caused by the satellite crossing the Southern Atlantic Anomaly (SAA). Due to the orbit geometry of the CoRoT satellite this event occurs quasi-periodic but is still obvious in the frequency analysis of most lightcurves as a period of 103 minutes ($\simeq0.162$~mHz or $\approx13.98$ cycles per day). 

Care must be taken for adjacent data points to marked ones, since the flagging might not always be appropriate. In fact the flagging is only dependent on the longitude and latitude of the satellite regardless of the size of the SAA. Random electronic events that alter the flux on a region of the CCD or the whole CCD are not flagged and must be detected and removed by the user (see Sec.~\ref{sec:jumps} for details).

The selection of valid data points can also be managed by the user himself in principle. Taking into account, that all lightcurves are affected in the same way, the outliers can be removed either by simple sigma clipping or by subtracting a median trend (cf. Sec.\ref{sec:detrending}).


\section{Data Reduction}
Possible contaminants, which are stars close to or even enclosed in the target PSF, should be checked in the \textsc{WinDescriptor} file or in the ExoDat database. Bright stars may also contaminate a complete pixel row or a pixel column (Cabrera, 2010). Stars in the same PSF can be either fainter background stars or other CoRoT targets nearby. As the degree of contamination is a function of distance and magnitude, the different color channels are mitigated unevenly. Contamination can also be identified by using a cross-correlation matrix of all lightcurves, where e.g. bright diluting stars form clusters in that matrix. 

The \textsc{SysRem} algorithm by Mazeh et al. (2009) minimizes signal cross-talk in a whole observation run and increases the overall signal-to-noise ratio. A similar technique for removing systematics in lightcurves has been presented by Tamuz (2005).  A much faster way is the determination and subtraction of a median lightcurve for the whole run. Even small subsets of lightcurves benefit from this procedure.


\subsection{Normalization}
The flux values are usually in the order of $3\cdot10^4$ to $2\cdot10^6$ and should be normalized to $1.0$ or converted to magnitudes for further numerical processing. This overcomes problems when fitting models to the lightcurve, since most of these algorithms are based on matrix inversion techniques which then cause singular values. Depending on the further application the data should be normalized by the median or the mean, which is also given in the FITS header as LC\_MEAN or LC\_MEANR (see Table~\ref{tab:modes}) for e.g. the red channel.


\subsection{De-trending}
\label{sec:detrending}
Almost all lightcurves suffer from a linear or polynomial trend which is either a stellar phenomenon e.g. long term variability or caused by the relativistic aberration due to the satellites orbit as it is co-moving with the earth around the sun. The latter can be identified as it affects the color channels differently and is usual evident in long runs as a sinusoid with a period of a sidereal year. On shorter runs this effect might not be visible or can be removed by the described polynomial.

\begin{figure}[ht]
\includegraphics[width=\columnwidth]{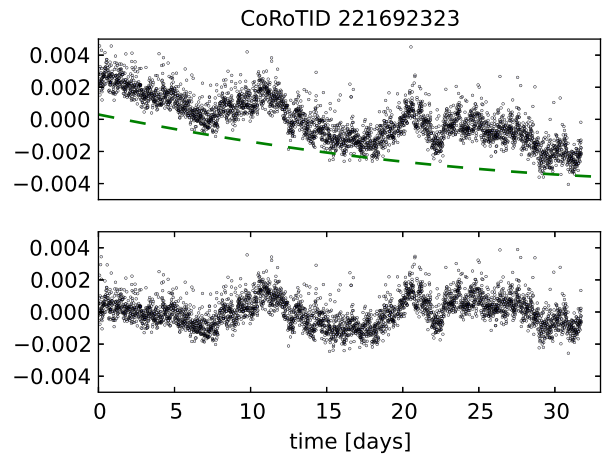}
\caption{The top plot shows the raw lightcurve with flagged data points removed and normalizing to $1.0$ and subtracting the mean value. The least squares fitted second order polynomial is plotted as a dashed line with an offset for clarity. After subtracting the fitted function a de-trended dataset (bottom plot) remains.}
\label{fig:detrend}
\end{figure}

When applying a polynomial fit to the data for detrending (e.g., De Medeiros et al., 2013 and references therein), the original measurement errors must not be taken into account. A better way would be fit with a smaller weight on outliers or an unweighted fit. This can be achieved in two stages by fitting a polynomial with uniform measurement-errors and afterwards taking the absolute difference between the flux and the polynomial as weights for the second fit (Figure~\ref{fig:detrend}).

It has to be considered that e.g. stellar rotation periods in the order of the run lengths, which are evident in the short observation runs of CoRoT, will be affected by this procedure. However, de-trending for short runs might not be necessary.


\subsection{Binning}
\label{sec:binning}
\begin{figure}[ht]
\includegraphics[width=\columnwidth]{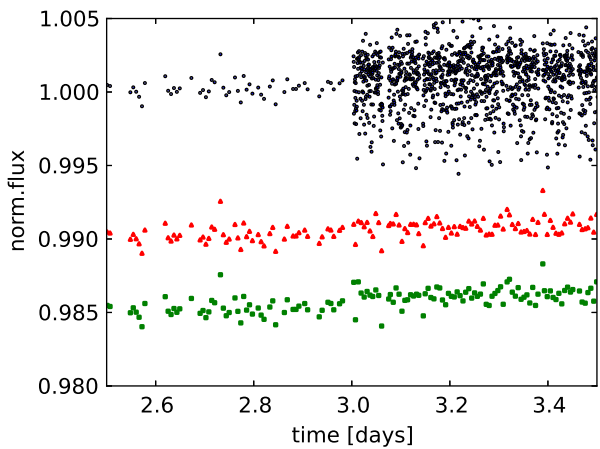}
\caption{The top part of the figure shows the original normalized lightcurve of CoRoTID~221694924 that was switched to a high cadence in alarm mode at day three after the beginning of the run. The middle lightcurve with an abitrary offset is the re-binned dataset by taking the mean value in each 512~second bin. The bottom lightcurve shows the re-binned dataset by taking the median in each bin, which modifies the statistics evidently by decreasing the noise level but also introducing a shift.}
\label{fig:rebin}
\end{figure}

The original data cadence on board the satellite is 32~seconds. To preserve the bandwidth limit and to increase the signal-to-noise ratio, the data is binned on-board to 512~seconds, which has to be considered for the epoch of the data points (cf. Sec.~\ref{sec:epoch}, and Auvergne et al. (2009)). Bright or interesting targets are observed at high cadence throughout the whole run. Targets that show transit-signatures in the early telemetry (so called \textit{alarm mode candidates}) are switched from 512~seconds to 32~seconds in high cadence mode (Quentin et al. 2006; Surace et al. 2008). This switch causes the noise-level to increase by a factor of 4 naturally, which is evident in the lightcurve as seen in Figure~\ref{fig:rebin}.

The on-board binning to 512~seconds generates the error estimates for the different channels. In lightcurves with mixed cadences or 32~second cadence this value is partly or totally missing. It is advisable to recalculate the error estimate after binning by using the rms or the mean-absolute-derivative.

One of the most accurate algorithms to re-sample the data is the one-dimensional application of the drizzle-algo\-rithm by Fruchter \& Hook (1997). By taking the median in each bin, the noise is reduced better than the on-board algorithm and the rms in each bin can provide a good error estimate(see Figure~\ref{fig:rebin}, bottom). If the noise-characteristics need to be preserved then the mean value in each bin has to be taken (see Figure~\ref{fig:rebin}, middle). This is advisable especially for alarm-mode candidates.


\subsection{Gaps and Jumps}
\label{sec:jumps}
By removing flagged or invalid data-points gaps will occur inevitably. These missing data-points create an obstacle for the Fast Fourier Transform (FFT) or other algorithms relying on equidistant data. To avoid this problem, the lightcurve can be reconstructed by applying a simple linear interpolation. Polynomial interpolation functions tend to induce spurious frequencies in the Fourier spectra of the lightcurves. In most applications data gaps are not necessarily to be filled since further processing steps might be phase binning or model fitting routines that do not require evenly spaced data. 

Single outliers can be removed by sigma clipping of the data points at e.g. $5\sigma$ (Debosscher et al., 2009) or their first derivative. The latter one is far more difficult, since the noise distribution is not uniform, which affects the first derivative. The first derivative can be evaluated using either the local differential quotients $\Delta f/\Delta t$ or by using a three point Lagrangian interpolation with its analytical derivative. Both methods seem feasible but handle noisy data differently. Ideally the first order derivative shows a normal (Gaussian) distribution. Interestingly for CoRoT data this distribution is not symmetrically and shows higher values at the positive end. 

Many of the lightcurves are degraded by one ore many jumps. They are caused by high energetic particles in the low Earth orbit of the spacecraft (see Pinheiro da Silva et al. 2008 for details). Alterations caused by electronic events are difficult to detect and to handle. Large spikes or drops often occur on several lightcurves concurrently and can be treated easily by either removing the affected data points or by sigma clipping. The Savitzky-Golay-Filter (Savitzky \& Golay, 1964), which is often applied according to literature tends to modify the signal in our case transits (see Barge et al. 2008, Alonso et al. 2008, Rauer et al. 2009, Fridlund et al. 2010). Its filtering window has to be selected carefully and is usually chosen to be 179.2~minutes long, which is well below one day, but longer than CoRoT's orbital period.

The method provided by Mislis et al. (2010) is not satisfactory, since it only works for lightcurves with three color-channels. Other methods are semi-automatic and include visual inspection (De Medeiros et al., 2013). Again jumps can be detected in principle by analysing the first derivative of the lightcurve.

Jumps can be removed by applying a $1\sigma$ to $3\sigma$ clipping to the first derivative. After that an integration or a cumulative sum $f_i = \sum_{j=0}^i \Delta f_j/\Delta t_j$ of the finite differences produces a filtered signal. However, since many jumps alter the statistics of the lightcurve and its standard deviation the sigma limit for the clipping is an individual choice for each lightcurve and can not be set globally. Stricter limits tend to alter intrinsic signals like eclipses or rotational modulation, were on the other hand a high limit e.g. $5\sigma$ might usually miss the jumps.


\subsection{Frequency filtering}
\begin{figure}[ht]
\includegraphics[width=\columnwidth]{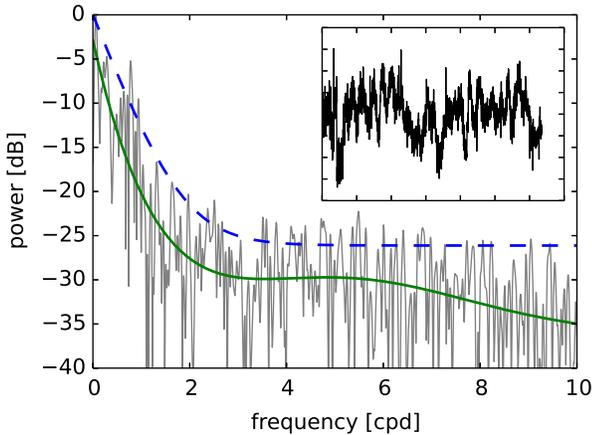}
\caption{The power spectrum of CoRoTID 221622012 is shown. The green solid line shows the exponential polynomial fit as described in the text and the blue dashed line represents the fit with $a \cdot exp(-b \cdot f) + c$. The data has been zero-padded by the same length of the dataset and a Hamming window has been applied before calculating the Fourier spectrum. The inset shows the lightcurve after de-trending for comparison.}
\label{fig:rednoise}
\end{figure}

Principally all lightcurves are affected by stellar red noise in the frequency domain. A simple approximation of the shape of the power spectrum has been demonstrated by Harvey et al. (1993) with 
\begin{equation}
P(\nu) = A/[1+(2\pi\nu p_0)^b],
\end{equation}
where $A$ is the amplitude, $p_0$ is the characteristic time scale and $-2/b$ the decay rate. An example of red noise can be seen in Figure~\ref{fig:rednoise}. 
The red noise component can be mostly suppressed in the frequency domain between 0.04\,cpd and 11\,cpd by dividing or subtracting an exponential polynomial fit $\exp(P(f,5))$ of the Fourier power spectrum, where $P(f,5)$ is a fifth order polynomial as a function of the frequency. This procedure flattens the noise background, improving the detection of periodic signals, since single peaks are apparent over the background. This can be seen in Figure~\ref{fig:rednoise} for the peak at approximately one day (here 1~cpd), which is well above the green solid line, while almost vanishing when using the lower exponential polynomial fit represented by the blue dashed line. An application of this method on lightcurves for purposes of increasing the signal to noise is not advisable, since it might produce unrealistic results.

\begin{figure}[ht]
\includegraphics[width=\columnwidth]{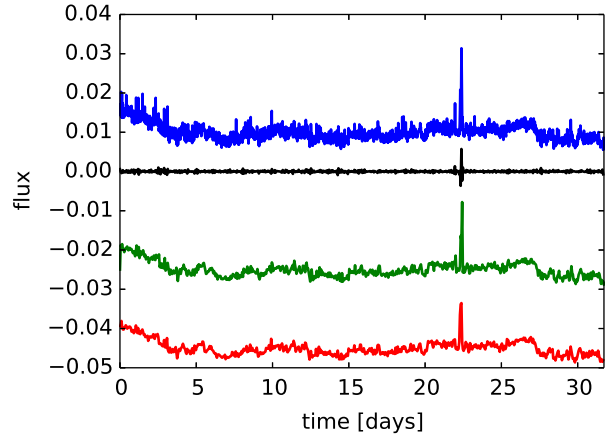}
\caption{Resampling as demonstrated on CoRoTID 221678744. The top graph represents the original lightcurve with an arbitrary offset, where as the residuals are shown in black at zero offset with the correct scaling. The green lightcurve below has been downsampled by a factor of 8 using a Fourier cut-off and the lowest lightcurve shows the rebinned version by the same factor. Except for the outlier the two reduced lightcurves are identical within 0.038\%.}
\label{fig:resamp}
\end{figure}

Red noise, signals of stellar rotation or other astrophysical phenomena inhibit the application of a Wiener-filter to remove unwanted noise. Nevertheless one might apply a bandpass- or a low-pass filter in order to investigate the relevant frequency-domain. A simple frequency cut-off is equivalent to a resampled lightcurve (see Sec.~\ref{sec:binning}), which might be sufficient and computationally advantageous (see Figure~\ref{fig:resamp}). 

One should consider to remove the signal imposed by CoRoT's orbital period and its harmonics to avoid confusion in further processing steps. While the usage of flagged data might increase the number of usable datapoints, the amplitude of the satellite's orbit in the power-spectrum might also increase. A notch-filter can be applied to 13.972~cpd ($\approx 6184$~seconds, see Auvergne et al. (2009)) and its harmonics, to remove the respective frequency from the power spectrum. A different way to remove the imprint of CoRoT's orbital period is the harmonic filtering described by Grziwa et al. (2012).

\begin{figure}[ht]
\includegraphics[width=\columnwidth]{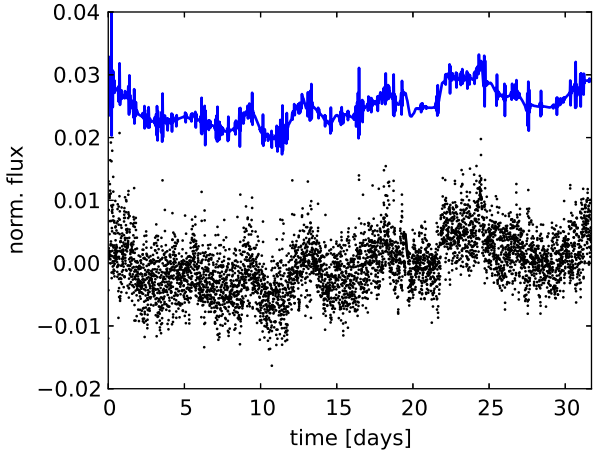}
\caption{The figure above shows the demonstration of the wavelet filter applied on the lightcurve of CoRoTID 221652737. The lower scatter plot shows the original lightcurve after normalization, re-binning and de-trending. The upper line shows the filtered lightcurve using the Daubechies-Wavelet 'db4', a symmetric extension with 80 coefficients and soft thresholding. }
\label{fig:wave}
\end{figure}

A more sophisticated approach is the cleaning of lightcurves by applying a wavelet filter as demonstrated in Figure~\ref{fig:wave}.

Best results are achieved by using the Daubechies-Wavelet\footnote{\url{http://code.google.com/p/wavepy/}} using between 75 to 100 coefficients. The selection of wavelet as well as the decision between a soft and a hard thresholding has to be based on the kind of application.

\section{Data Analysis}

Obviously the type of data analysis strongly depends on the science goal that has to be achieved. For a homogeneous analysis, it is recommendable to treat all observing runs in a equal manner (De Medeiros et al., 2013; Debosscher et al., 2009). Global corrections to the data might be applied in a hierarchical manner, on the level of runs, CCDs (E1, E2) or the amplifiers (L, R).

For the detection of exoplanets many different algorithms are in use and one is referred to the work of Cabrera et al. (2012) and references therein.

For the analysis of periods and frequencies the development for algorithms like the Lomb-Scargle periodogram (Lomb, 1976; Scargle, 1981) like was driven by ground based data suffering from gaps caused by the day-night cycle. These problems are less of an issue in space photometry data, which only contain minor gaps, as evident in the CoRoT data.

\begin{figure}[ht]
\includegraphics[width=\columnwidth]{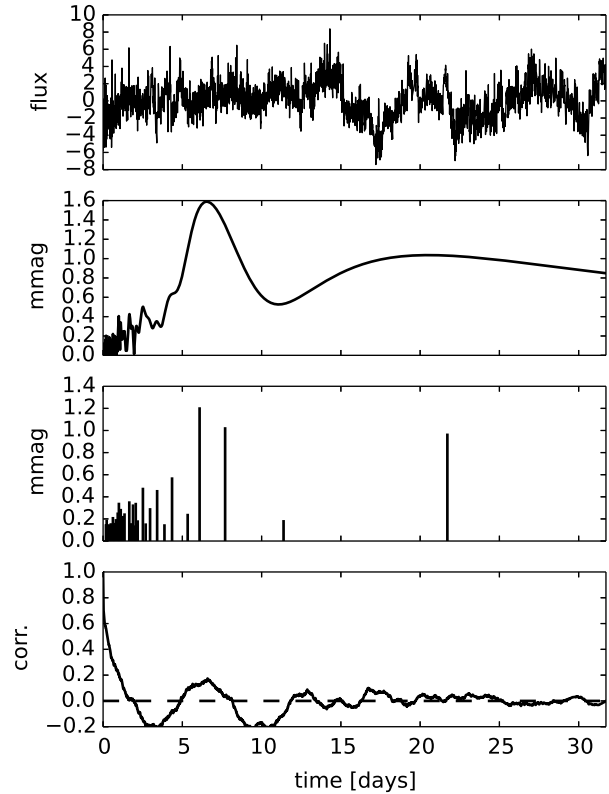}
\caption{The lightcurve of CoRoTID~221665817 (top panel) has been detrended, rebinned and the gaps have been interpolated and is analysed by different methods. The second panel shows the amplitude spectrum as calculated with FFT. The third panel shows the harmonic fitting of the first 30 characteristic frequencies. The lower panel shows the autocorrelation function. The flux has been scaled by a factor of 1000.}
\label{fig:freqanalysis}
\end{figure}

Nevertheless the Lomb-Scargle periodogram is commonly used in literature to perform the analysis in frequency space. Though it is a good substitute for the FFT, and does not rely on equidistant data points, there are better and more elegant solutions. One of them is the (plain) least squares periodogram described by Zechmeister \& K\"{u}rster (2009) or a similar method by Plavchan et al. (2008). One shall be warned that these algorithms are not optimized for large datasets, hence computationally intensive. So filling the missing datapoints as described in Section~\ref{sec:jumps} and applying a FFT is a good trade-off between fast processing speed and high accuracy. The different period analysis methods can be seen in Figure~\ref{fig:freqanalysis}.

Another alternative for detecting periodic signals is the autocorrelation function (ACF) as described in McQuillan et al. (2013). The ACF provides a stable method to determine the period in a regime typical between half a day and half the length of the run. A low SNR requires additional measures to be taken, e.g. by folding the ACF with the Fourier periodogram to obtain the correct peak.


\section{Conclusions}
The CoRoT data has been very carefully treated in the N0 pipeline to preserve the scientific contents in the public available N2 data product. This leaves a major part of the data preparation to the user.

Now that all observing runs are public, a homogeneous processing is possible but at the cost of large computing power.
 
Certainly, the choice of data reduction tools strongly depends on the scientific case, but most of the time a detrended lightcurve and an equally spaced time series is a minimum requirement to analyse the data further. The main issues of the CoRoT data, namely long term flux variations (trends) and spikes and jumps can be removed to some extend in an automated way leaving the fine-tuning and cleaning of some complicated cases to the human analyst.

Over all using the set of tools described explicitly in this paper or the references to alternative methods should guide the user of the CoRoT data product through the data reduction process.

\acknowledgements
I want to thank the anonymous referee for his valuable comments and suggestions improving the overall quality of this publication. This research has made use of the ExoDat Database, operated at LAM-OAMP, Marseille, France, on behalf of the CoRoT/Exoplanet program.

\end{document}